\documentclass[12pt]{article}
\usepackage{epsfig}
\usepackage{latexsym}
\usepackage{color}
\newcommand{\mysquare}[0]{\raise-.2ex\hbox{{\Large$\Box$}}}
\def\lsim{\mathrel{\rlap {\raise.5ex\hbox{$ < $}}
{\lower.5ex\hbox{$\sim$}}}}
\def\gsim{\mathrel{\rlap {\raise.5ex\hbox{$ > $}}
{\lower.5ex\hbox{$\sim$}}}} \topmargin -1.5cm \textheight=22.5cm
\catcode`\@=11
\newcount\hour
\newcount\minute
\newtoks\amorpm
\hour=\time\divide\hour by60
\minute=\time{\multiply\hour by60 \global\advance\minute by-\hour}
\edef\standardtime{{\ifnum\hour<12 \global\amorpm={am}%
        \else\global\amorpm={pm}\advance\hour by-12 \fi
        \ifnum\hour=0 \hour=12 \fi
        \number\hour:\ifnum\minute<10 0\fi\number\minute\the\amorpm}}
\edef\militarytime{\number\hour:\ifnum\minute<10 0\fi\number\minute}
\def\draftlabel#1{{\@bsphack\if@filesw {\let\thepage\relax
   \xdef\@gtempa{\write\@auxout{\string
      \newlabel{#1}{{\@currentlabel}{\thepage}}}}}\@gtempa
   \if@nobreak \ifvmode\nobreak\fi\fi\fi\@esphack}
        \gdef\@eqnlabel{#1}}
\def\@eqnlabel{}
\def\@vacuum{}
\def\draftmarginnote#1{\marginpar{\raggedright\scriptsize\tt#1}}
\def\draft{\oddsidemargin -.2truein
        \def\@oddfoot{\sl preliminary draft \hfil
        \rm\thepage\hfil\sl\today\quad\militarytime}
        \let\@evenfoot\@oddfoot \overfullrule 3pt
        \let\label=\draftlabel
        \let\marginnote=\draftmarginnote
   \def\@eqnnum{(\theequation)\rlap{\kern\marginparsep\tt\@eqnlabel}%
\global\let\@eqnlabel\@vacuum}  }

\relax

\newcommand{\ba}[0]{\begin{eqnarray}}
\newcommand{\ea}[0]{\end{eqnarray}}
\def\bs{\begin{subequations}}
\def\es{\end{subequations}}

\def\thebibliography#1{%
\vskip 0.5cm \centerline{\bf References}
\list{%
[\arabic{enumi}]}{\settowidth\labelwidth{[#1]}
\leftmargin\labelwidth
\advance\leftmargin\labelsep
\usecounter{enumi}}
\def\newblock{\hskip .11em plus .33em minus .07em}
\sloppy\clubpenalty4000\widowpenalty4000
\sfcode`\.=1000\relax}

\renewcommand{\theequation}{\arabic{section}.\arabic{equation}}

\renewcommand{\section}{\setcounter{equation}{0}\@startsection%
{section}{1}{0mm}{-\baselineskip}{0.5\baselineskip}%
{\normalfont\normalsize\bfseries}}

\renewcommand{\subsection}[1]{\addtocounter{subsection}{1}
\vspace{2.5mm}\par\noindent {\it \thesubsection . #1}\par
 \vspace{0.5mm} }

\renewcommand{\subsubsection}{\@startsection%
{subsubsection}{3}{0mm}{-\baselineskip}{0.5\baselineskip}%
{\normalfont\normalsize\slshape}}

\usepackage{amscd} \usepackage{amsmath} \usepackage{amsfonts}

\def\s{\sigma}

\def\thefootnote{\fnsymbol{footnote}}
\def\es{\end{subequations}}

\def\ec{\hat E^{c}_{2}}

\newcommand{\uarrw}[0]{\mathrel{
{\raise.5ex\vbox{\hrule width 1cm}\hskip-6pt\rightarrow}}}

\def\bea{\begin{array}}
\def\bem{\begin{displaymath}}
\def\beq{\begin{equation}}

\def\eea{\end{array}}
\def\eem{\end{displaymath}}
\def\eeq{\end{equation}}

\def\s2w{\sin^2 \theta_W}
\def\Tr{\mathop{\rm Tr}}

\def\be{\begin{equation}}
\def\ee{\end{equation}}
\def\bc{\begin{center}}
\def\ec{\end{center}}
\def\bea{\begin{eqnarray}}
\def\eea{\end{eqnarray}}

\begin{document}
\begin{titlepage}
\begin{flushright}
LPTENS--08/43
\\
CPHT--RR055.0708
\\
DFTT 24/2008
\end{flushright}

\vspace{2mm}

\begin{centering}
{\bf\large  RESOLUTION OF HAGEDORN SINGULARITY }\\
\vspace{10pt}
{\bf\large  IN SUPERSTRINGS WITH GRAVITO-MAGNETIC FLUXES}\\

\vspace{20pt}
 {Carlo Angelantonj$^{1}$, Costas~Kounnas$^{2}$,
 Herv\'e~Partouche$^3$ and Nicolaos~Toumbas$^4$}

\vspace{.3cm}

$^1$ Dipartimento di Fisica Teorica and INFN Sezione di Torino\\
Via  P. Giuria 1, IÐ10125 Torino, Italy.
{\em  carlo.angelantonj@cern.ch}

\vskip .1cm

$^2$ Laboratoire de Physique Th\'eorique,
Ecole Normale Sup\'erieure,$^\dagger$ \\
24 rue Lhomond, F--75231 Paris Cedex 05, France.
{\em  kounnas@lpt.ens.fr}

\vskip .1cm

$^3$ Centre de Physique Th\'eorique, Ecole Polytechnique,$^\diamond$
\\
F--91128 Palaiseau, France.
{\em partouche@cpht.polytechnique.fr} \vskip .1cm

$^4$ Department of Physics, University of Cyprus,\\ Nicosia 1678, Cyprus. {\em nick@ucy.ac.cy}
 \vspace{6mm}

{\bf Abstract}

\end{centering}

\noindent 
We consider closed type II and orientifold backgrounds where supersymmetry is spontaneously broken by asymmetric geometrical fluxes. We show that these can be used to describe thermal ensembles with chemical potentials associated to ``gravito-magnetic'' fluxes. The thermal free energy is computed 
at the one-loop string level, and it is shown to be free of the usual Hagedorn-like instabilities for a certain
choice of the chemical potentials. In the closed string gravitational sector, as well as in the open string matter sector of the proposed orientifold construction, the free energy turns out to have ``Temperature duality'' symmetry, ${\cal F}(T/T_H)={T^2\over T_H^2}~{\cal F}(T_H/T)$, which requires interchanging the space-time spinor representations $S\leftrightarrow C$. For small temperatures, $T\to 0$,  the anti-spinor $C$ decouples from the spectrum while for large temperatures, $T\to \infty$,  the spinor $S$ decouples. In both limits the free energy vanishes, as we recover a conventional type II superstring theory. At the self dual point $T=T_H$, the thermal spectra of $S$ and $C$ are identical. Moreover,
there are extra massless scalars in the adjoint representation 
of an $SO(4)$ non-abelian gauge symmetry in the closed-string sector, and open-string massless states charged simultaneously under both the Chan-Paton and the closed-string $SO(4)$ gauge group.

\vspace{5pt} \vfill \hrule width 6.7cm \vskip.2mm{\small \small \small \noindent
$^\dagger$\ Unit{\'e} mixte  du CNRS et de l'Ecole Normale Sup{\'e}rieure associ\'ee \`a
l'Universit\'e Pierre et Marie Curie (Paris VI), UMR 8549.\\
 $^\diamond$\ Unit{\'e} mixte du CNRS et de l'Ecole Polytechnique,
UMR 7644.}

\end{titlepage}
\newpage

\setcounter{footnote}{0}
\renewcommand{\thefootnote}{\arabic{footnote}}

\setlength{\baselineskip}{.7cm}
\setlength{\parskip}{.2cm}

\setcounter{section}{0}


\section{Introduction}

The thermal partition function of a (supersymmetric) field theoretic system can be obtained 
via a Euclidean path integral, where the Euclidean time is compactified on an $S^1$ cycle with period inversely proportional to the temperature: $2\pi R= T^{-1}$. Bosonic fields obey periodic boundary conditions along this cycle while fermionic fields obey 
anti-periodic boundary conditions. Equivalently, the thermal partition function can be derived by summing over particle paths in a first quantized approach. The spin-statistics connection requires that each path be weighted by the phase $(-1)^{\tilde {m}F}$, where the integer $\tilde{m}$ counts the number of times the path winds the $S^1$ cycle, and $F$ is the space-time fermion number. As a result, supersymmetry is always broken in the effective thermal theory. 
 
This construction can be generalized to superstring theories 
\cite{AtickWitten,RostKounnas,Kutasov:1990sv,AKADK}. Here, 
one encounters new phenomena due to the extended nature of strings and due to the presence of both momentum and winding states. 
For closed strings, the world-sheet degrees of freedom split into left-moving 
and right-moving excitations, both contributing to the space-time fermion number: 
$F= F_{ L} + F_{R}$.  We denote by $F_{L}$ the contribution of the world-sheet left-movers to the space-time fermion number, and similarly for $F_{ R}$.  In the heterotic string, $F_{R}$ is always even. $F_{ L}$ is even in the Neveu-Schwarz (NS) sector 
and odd in the Ramond (R) sector. Therefore the spin-statistics properties of space-time particles are determined only by the left-moving world-sheet degrees of freedom.    
In the type II superstrings on the other hand, both $F_{ L}$ and $F_{R}$ can be even or odd. 
In fact, these theories admit two independent $Z_2$ symmetries, $(-1)^{F_{ R}}$ and $(-1)^{F_{ L}}$,  under which the right-moving and left-moving R sectors change sign, respectively.

In the heterotic string, the thermal partition function is obtained by the temperature 
phase insertion \cite{AtickWitten,RostKounnas}
\begin{equation}
  (-1)^{\tilde {m}F_{ L}+n\tilde F_{ L}+\tilde{m}n }.
\end{equation}
This definition is indeed unique, dictated by the spin-statistics connection and 
modular invariance. In the type II closed strings the thermal phase insertion 
\begin{equation}
 (-1)^{\tilde m (F_{ L}+F_{ R})+n(\tilde F_{ L}+\tilde F_{ R}) }
\end{equation}
breaks the initial ${\cal N} \le 8$ supersymmetry giving rise to a non trivial free energy density 
similar to the heterotic string case.
In the type II case, the free energy is well defined as long as 
\be
T \ll T_H,~~~~ R_0 \gg R_H = { \sqrt{2\alpha^\prime}},
\ee  
where $T_H$ is the Hagedorn temperature \cite{Hagedorn,AtickWitten,RostKounnas,Kutasov:1990sv,AKADK}. (In any given thermal string model, the Hagedorn temperature corresponds to the temperature at which a certain string mode winding along the Euclidean time circle becomes massless). 
In the heterotic string, although there are tachyonic instabilities beyond the Hagedorn temperature, thanks to ``Temperature duality'',
$T \to 1/T$, there is a phase transition to a stable vacuum characterized by a condensation of thermal winding states \cite{AKADK}. 
We would like to stress here that in the type II closed string theories, as well as in standard orientifold constructions \cite{ADS}, 
the canonical thermal system does not possess such a ``Temperature duality'' symmetry, 
and it was not known until now how to stabilize the high temperature phase, 
although some ideas to overcome this problem in lower dimensions have been presented in \cite{ACI}. 
As we show in this work, type II freely acting asymmetric orbifold constructions share similar self-duality properties 
with finite temperature heterotic strings, and moreover a stable vacuum can be found at the perturbative string level.  

In a very similar way to the thermal supersymmetry breaking, string models with spontaneous breaking of supersymmetry can be constructed, adapting the Scherk-Schwarz 
mechanism \cite{Scherk:1978ta} in 
superstrings \cite{Rohm,stringySS,RostKounnas,ShifftedLat,ADS,ACI,OpenSS}, which involves the introduction of non-trivial geometrical fluxes \cite{GeoFluxes}. 
These fluxes can be described naturally in the framework of freely acting orbifolds \cite{ShifftedLat}.
The breaking of supersymmetry via gravito-magnetic fluxes, associated to an 
$R$-symmetry charge $Q$, is achieved by the following phase insertion: 
\begin{equation}
 (-1)^{\tilde m (Q_{ L}+Q_{ R})+n(\tilde Q_{ L}+\tilde Q_{ R}) },
\end{equation} where we introduce a non trivial coupling of $Q$ to the winding numbers $(\tilde m, n)$ along a spatial $S^1$ cycle. 
As in the temperature breaking case, here also tachyonic instabilities arise when the radius of the
compact spatial cycle is smaller than the Hagedorn radius.  However, when both thermal and flux breakings are present, a vast range of possibilities exists, which may or may not bypass the Hagedorn instabilities, as in \cite{ACI}.  

In this work we will present type II closed string and orientifold models, where supersymmetry is spontaneously broken 
by asymmetric geometrical fluxes along a two-dimensional space-time torus. One cycle of the torus is taken to be the compact Euclidean time direction. 
As we will show, these models admit a thermal interpretation, where additional sources
of spontaneous supersymmetry breaking in the form of gravito-magnetic fluxes 
are turned on.
A very interesting result is that the usual Hagedorn-like instabilities are absent 
for certain choices of chemical potentials associated to the gravito-magnetic fluxes.
As the resulting spectrum is free of tachyonic excitations, the free energy is finite for any  value of the temperature.

The structure of the paper is as follows. In Section 2 we present the tachyon free, asymmetric type II models and discuss their thermal interpretation. 
Section 3 is devoted to the orientifold construction, where the open string matter is
introduced. The free energy turns out to possess a novel ``Temperature duality symmetry", ${\cal F}(T/T_H)={T^2 \over T_H^2}~{\cal F}(T_H/T)$, in both the closed-string gravitational sector as well as in the open-string matter sector of the orientifold. 
Fermions are massive for any value of $T$. At the self dual point, $T=T_H$, extra massless scalars emerge, 
transforming in the adjoint representation of an $SO(4)$ non-abelian gauge symmetry group, 
originating from the close-string sector.  
Finally, in Section 4 we draw our conclusions and speculate about 
possible cosmological consequences due to the absence of a Hagedorn transition in this class of models.

\section{The closed string sector}

Our starting point is the type IIB superstring compactified on a two-torus with coordinates $(x^0,x^1)$. Depending on the context, these coordinates label two internal (compact) space-like directions or, alternatively, the Euclidean time $x^0$ and an additional space-like direction \cite{Cosmo-RT-Shifted,  Mod-Stab}. The first case can be used to describe string compactifications where supersymmetry is spontaneously broken by geometrical fluxes (e.g. stringy Scherk-Schwarz compactifications) \cite{stringySS,RostKounnas,ShifftedLat,OpenSS},  with the internal $T^2$ moduli corresponding to physical, fluctuating fields \cite{Mod-Stab}.
The second case is suitable to describe statistical, thermal-like ensembles \cite{Cosmo-RT-Shifted, Mod-Stab}. 
As we will see, in this second case the Euclidean $T^2$ geometrical data 
appear as thermodynamical parameters of the thermal ensemble. 
In particular, they do not give rise to any fluctuating fields in the low-energy description 
since local worldsheet symmetries allow us to gauge away all oscillator modes in the 
$X^\pm \sim X^1 \pm i X^0$ string coordinates.

The initially supersymmetric partition function is given by
\begin{equation}
{\cal Z} = {\textstyle\frac{1}{4}} \int_{\cal F} \frac{d^2 \tau}{\tau_2} \, 
\frac{1}{(\eta \bar \eta )^{12}}\,\Gamma_{(2,2)}\, \Gamma_{(8,8)} \,  \sum_{a,b=0,1} (-1)^{a+b+ab} \, \theta^4 \left[ {a \atop b}\right]
\, \sum_{\bar a , \bar b =0,1} (-1)^{\bar a + \bar b + \bar a \bar b} \, \bar\theta^4 \left[ {\bar a \atop \bar b}\right] \,,
\label{partition}
\end{equation}
where the two-dimensional $\Gamma_{(2,2)}$ lattice describes the contribution of the zero-modes of the $(x^0,x^1)$ coordinates, 
while the $\Gamma_{(8,8)}$ lattice describes the contribution of the zero-modes of the 
remaining coordinates. We compactify all ten Euclidean directions, 
regularizing the volume of space, in order to obtain a well defined, finite free energy in the cases for which supersymmetry is broken by thermal {\it and} quantum corrections \cite{Cosmo-RT-Shifted, Mod-Stab}. 

When the metric of the two-torus is diagonal, $G_{01}=0$,  and the antisymmetric 
tensor field $B_{01}$ vanishes, 
the $ \Gamma_{(2,2)} $ lattice factorizes in 
terms of two $\Gamma_{(1,1)}$ lattices,
$\Gamma_{(2,2)} = \Gamma_{(1,1)} (R_0)\, \Gamma_{(1,1)} (R_1)$, 
which are parameterized by the radii $R_0,~R_1$ of the corresponding circles, respectively. The $\Gamma_{(1,1)}$ lattice
is given by
\begin{equation}
\Gamma_{(1,1)} (R) = ~\frac{R}{\sqrt{\tau_2}}\sum_{\tilde m , n} e^{-\pi \frac{R^2}{\tau_2} |\tilde m + n \tau |^2}
=~ \sum_{m,n} \Gamma_{m,n} \,,             
\label{lattice}
\end{equation}
where 
\begin{equation}
\Gamma_{m,n} = q^{\frac{1}{4} p_{ L}^2 }\, \bar q ^{\frac{1}{4} p_{ R}^2}\;,\quad   q=e^{2\pi i \tau}\;,\quad \mbox{and}\quad      p_{ L,R} = \frac{m}{R} \pm n R\,
\end{equation}
are the left-moving and right-moving momenta. The expression in terms of $~p_{ L}$ and $p_{ R}$ is obtained as usual via a Poisson re-summation over the winding number $\tilde m$.

For later use, it is convenient to express the partition function (\ref{partition})  in terms of the $SO(8)$ characters,
\begin{equation}
O_8 = \frac{\theta_3^4 + \theta_4^4 }{ 2\,\eta^4}\,,
\qquad 
V_8 = \frac{\theta_3^4 - \theta_4^4}{ 2\, \eta^4}\,,
\qquad
S_8  = \frac{\theta_2^4 - \theta_1^4}{ 2\,\eta^4}\,,
\qquad
C_8= \frac{\theta_2^4 + \theta_1^4 }{2\, \eta^4}\,,
\label{characters}
\end{equation}
as follows:
\begin{equation}
{\cal Z} = \int_{\cal F} \frac{d^2 \tau}{\tau_2} \frac{1}{ (\eta \bar \eta )^8}\, |V_8 - S_8|^2 \, 
\Gamma_{(1,1)} (R_0)\, \Gamma_{(1,1)} (R_1 ) \, \Gamma_{(8,8)} \,.
\label{partham}
\end{equation}

Our main goal is to deform the initially supersymmetric theory, and in particular 
break space-time supersymmetry {\it asymmetrically} by coupling the $x^0$ lattice 
to the left-handed space-time fermion number $F_{ L}$ and the $x^1$ 
lattice to the right-handed space-time fermion number $F_{ R}$. 
Under $(-1)^{F_{ L}}$ ($(-1)^{F_{ R}}$), the left- (right-) moving R sector changes
sign. 
This property and genus-one modular invariance 
allow us to replace the $x^0$ and the $x^1$ lattices in the integrand of  (\ref{partham}) with  
\begin{equation}
\frac{R_0}{\sqrt{\tau_2}}\sum_{\tilde m_0,n_0} 
e^{-\frac{\pi R_0^2}{\tau_2}|\tilde m_0 +n_0\tau|^2} (-)^{\tilde m_0 a+n_0b+\tilde m_0 n_0}\,, \quad
\frac{R_1}{\sqrt{\tau_2}}\sum_{\tilde m_1,n_1} 
e^{-\frac{\pi R_1^2}{\tau_2}|\tilde m_1 +n_1\tau|^2} (-)^{\tilde m_1 \bar a+n_1\bar b+\tilde m_1 n_1}\,.
\label{thermalright}
\end{equation}
In this way, the $x^0$ lattice is ``thermally'' coupled to the left-moving world-sheet degrees of freedom, while the  $x^1$ lattice is ``thermally'' 
coupled to the right-moving world-sheet degrees of freedom.
Notice however that modular invariance does not fix uniquely the deformation but, as we discuss  later on, allows for the possibility of having non-trivial discrete torsion. 

After Poisson re-summations over the winding numbers $\tilde m_0$, $\tilde m_1$, we find the following shifts to the left-moving and right-moving momenta along the $x^0$ and $x^1$ circles:
\begin{equation}
p_{ L,R}^0=\frac{m_0-\frac{1}{ 2}(a+n_0)}{R_0}\pm n_0R_0,
~~~~
p_{ L,R}^1=\frac{m_1- \frac{1}{2}(\bar a+n_1)}{R_1}\pm n_1R_1 \,.
\label{shiftedmomenta}
\end{equation}
So the spectrum is deformed and space-time supersymmetry is spontaneously broken. 
In addition, the left GSO projection is reversed in the $n_0$-odd winding sector, 
while the right GSO projection is reversed in the $n_1$-odd winding sector
 \cite{AtickWitten,RostKounnas,Kutasov:1990sv,AKADK}.
The partition function takes the form
\begin{eqnarray}
{\cal Z} &=& \int_{\cal F} \frac{d^2 \tau}{\tau_2} \, \frac{\Gamma_{(8,8)}}{(\eta \bar \eta)^8}  \, \sum_{m_0, n_0}\left(
V_8 \, \Gamma_{m_0,2n_0} + O_8 \, \Gamma_{m_0 +\frac{1}{2},2n_0 +1} - S_8 \, \Gamma_{m_0 +\frac{1}{2} , 2 n_0} -C_8\, \Gamma_{m_0, 2n_0 +1} \right)
\nonumber\\
& & \times \sum_{m_1,n_1}\left(
\bar V_8\, \Gamma_{m_1,2n_1} + \bar O_8 \, \Gamma_{m_1 +\frac{1}{2},2n_1 +1} - \bar S_8 \, \Gamma_{m_1 +\frac{1}{2} , 2 n_1} -\bar C_8\, \Gamma_{m_1, 2n_1 +1} \right) \,.
\label{part}
\end{eqnarray}

As a result, the only massless states emerge in the $ V \bar V$ sector that comprises the (reduction of) the ten-dimensional metric, Kalb-Ramond field and dilaton field. 
The initially massless fermions get a mass inversely proportional to the compactification radii
\begin{equation}
2 \, m^2_{{ V}{\! \bar S}} = \frac{1}{(\sqrt{2} R_1)^2} \,,
\qquad  2 \, m^2_{{ S}{ \bar V}} = \frac{1}{(\sqrt{2} R_0)^2}\,. 
\label{massS}
\end{equation}
Fermions arise also from the ${ V} { \bar C}$ and ${ C}{\bar V}$ sectors, which carry non-zero
winding charges, and their lightest masses are given by  
\begin{equation}
2 \, m^2_{{V}{ \!\bar C}} = (\sqrt{2} R_1)^2 \,, \qquad 2 \, m^2_{{ C}{\bar V}} = (\sqrt{2} R_0)^2\,.
\label{massC}
\end{equation}

For this asymmetric model the RR sectors are also lifted, with masses 
$$
~~~~~~~~~~~~2 \, m^2_{{ S} { \bar S}} =  {1\over (\sqrt{2} R_0)^2} + {1\over (\sqrt{2} R_1 )^2}
\,, \qquad 
2\, m^2_{{ C} { \bar C}} = (\sqrt{2} R_0)^2 + (\sqrt{2} R_1 )^2
\,,~~~~~~~~~
$$
\be
2\, m^2_{{ S} {  \bar C}}=   {1\over (\sqrt{2} R_0)^2} + (\sqrt{2} R_1 )^2
\,, \qquad 
 2\, m^2_{{ C} { \bar S}} = (\sqrt{2} R_0)^2 + {1\over(\sqrt{2} R_1 )^2}
\,.
\label{RR}
\ee
The reason is that the RR fields are charged under both the $Z_2$ symmetries $(-1)^{F_{ L}}$ and $(-1)^{F_{ R}}$. Thus the asymmetric breaking of supersymmetry also leads
to a spontaneous breaking of the ${ U}(1)$ gauge symmetries associated to the RR fields. 

Particularly interesting is the ${ O} { \bar O}$ sector. It includes the NS-NS vacuum, which typically in other models becomes tachyonic in some regions of the moduli space. When this happens, the free energy diverges and the system undergoes a first-order phase transition. 
In our case, however, the NS-NS vacuum always carries non-vanishing momentum and winding excitations, and its lightest mass is given by
\begin{eqnarray}
2 \, m^2_{{ O} { \bar O}} &=& \left( \frac{1}{2 R_0^2} + 2 R_0^2 + \frac{1}{2 R_1^2} + 2R_1^2 \right) - 4
\nonumber
\\
&=& \left( \frac{1}{\sqrt{2} R_0} - \sqrt{2} R_0 \right)^2 + \left( \frac{1}{\sqrt{2} R_1} - \sqrt{2} R_1 \right)^2 \,.
\label{tachyonmass}
\end{eqnarray}
The mass is always positive, except for the ``fermionic point" at radii $R_0 =R_1= 1/\sqrt{2}$, 
where new massless states emerge. 

In the infinite radii limit, $R_0$, $R_1 \to \infty$, we recover the massless chiral spectrum of the supersymmetric type IIB theory:
$m^2_{{ V}{ \!\bar S}},~ m^2_{{ S}{ \bar V}},~ m^2_{ S \bar S} \to 0$. All other masses listed  
above become infinite in this limit, and so the corresponding states decouple from the spectrum.
In the small radii limit, $R_0,R_1 \to 0$, we get the massless chiral spectrum of the 
(equivalent) supersymmetric type IIB$^\prime$ theory \cite{Polchinski,review}:
$m^2_{ V\!\bar C}, ~m^2_{C \bar V}, ~m^2_{  C\bar C} \to 0$.  
In fact by two T-duality transformations, one along the $x^0$ cycle
and one along the $x^1$ cycle, it is easy to see that the model at radii
$(R_0,R_1)$ is equivalent to the asymmetric type IIB$^\prime$ model 
at radii $({1 / 2R_0}, {1/ 2R_1})$. 
Under T-duality transformations, we have now that $\sqrt{2} R \to 1/\sqrt{2} R$,  and as usual the 
${S}$ and ${ C}$ sectors get interchanged.  At the boundaries of moduli space, $R_0$, $R_1$  $\to 0$ or $ \infty$,  the ${ O}$ sector together with either the $S$ or the $C$ Ramond sectors become infinitely massive and half or all of supersymmetries are recovered. 

The appearance of extra massless states at the fermionic point $\sqrt{2}R_0=\sqrt{2}R_1=1$ is a signal of enhanced gauge symmetry. 
This point is self-dual under the two T-dualities we mentioned above. 
At this point, the sectors ${V\bar O}$ and ${ O\bar V}$ can be level matched, giving rise to extra massless gauge bosons. 
The underlying reason is extended symmetry on the world-sheet.
When both radii are at the fermionic radius,
each of the left-moving bosons $X^0_L$ and $X^1_L$ is equivalent to two left-moving real fermions. Together with
$\psi^0$ and $\psi^1$, these generate an $ SO(4)_L\sim SU(2)_{2}\times SU(2)_{2}$ current algebra on the left side of the 
world-sheet \cite{ABKW,ABK,KF,massiveSUSY}. Similarly we have a right-moving $ SO(4)_R\sim SU(2)_{2}\times SU(2)_{2}$ current algebra.
Consequently, the spectrum includes non-abelian gauge bosons in the adjoint of the $SO(4)_L \times SO(4)_R$ gauge group. 

\subsection{Thermal interpretation}

In order to exhibit the thermal nature of the deformation, it is convenient to shift 
the winding numbers associated to the $x^1$ cycle as follows:
\be
\tilde m_1 \to \tilde m_1 + \tilde m_0 \,, \qquad n_1 \to n_1 + n_0\,.
\ee
The $T^2$ lattice takes now the form
\begin{eqnarray}
&&
\frac{R_0 \, R_1}{\tau_2} \sum_{\tilde m_0 , n_0} \sum_{\tilde m_1 , n_1}  e^{-\frac{\pi}{\tau_2}\, \left[ R_0^2 |\tilde m _0 + \tau n_0 |^2 + R_1^2 |\tilde m_1 + G \tilde m_0 +\tau (n_1 + G n_0 )|^2 \right]} \, e^{2 i \pi B (\tilde m_1 n_0 - \tilde m_0 n_1 )}
\nonumber \\
&&\qquad\qquad\qquad \times (-1)^{\tilde m_0 (a+\bar a ) + n_0 (b+\bar b)}\, (-1)^{\tilde m_1 \bar a + n_1 \bar b + \tilde m_1 n_1} \,.
\label{newlattice}
\end{eqnarray}
It is then clear, that along the $x^0$ cycle the deformation acts as a standard temperature deformation, since the $x^0$ cycle is coupled to the total fermion-number operator $F=F_{ L} + F_{ R}$. The  $x^1$ cycle couples only to the right-moving fermion number $F_{ R}$. 
The latter deformation breaks the initial ${\cal N}=(4,4)$ supersymmetries 
 to ${\cal N}=(4,0)$, which in turn are broken by thermal effects. 
Notice that in this representation the metric of the two-dimensional torus is not diagonal and there is a non-trivial $B$-field background
\be
ds^2 = R_0^2 (dx^0)^2 + R_1^2 ( dx^1 + G \, dx^0 )^2 \,,
\qquad
B_{ab} = \left( \begin{matrix} 0 & B \\ - B & 0 \end{matrix} \right).
\ee
Our model corresponds to the special values $G=2B=1$, 
although one may consider other generic values (see also the discussion at the end of this section). 
After a Poisson re-summation over $\tilde m_1$, one obtains
\begin{eqnarray}
&& \frac{R_0}{\sqrt{\tau_2}} \, \sum_{\tilde m_0 , n_0 } e^{-\frac{\pi R_0^2}{\tau_2} |\tilde m_0 + \tau n_0 |^2} \, (-1)^{\tilde m_0 (a+\bar a) +  n_0 (b+\bar b)}
\nonumber \\
&& \times \sum_{m_1 , n_1} q^{\frac{1}{4} p_{ L}^2 } \, \bar q ^{\frac{1}{4} p_{ R}^2}\, 
(-1)^{n_1 \bar b}\, e^{-i\pi \tilde m_0 \left[ 2B (n_1 + G n_0) - 2G m_1 +  G (\bar a + n_1)\right]}
\label{resummed}
\end{eqnarray}
where the left-moving and right-moving momenta associated to the $x^1$ cycle are given by
\be
p_{ L,R} = \frac{m_1 - \frac{1}{2} \, (\bar a + n_1 + 2 B n_0)}{R_1} \pm (n_1 + G n_0 ) R_1 \,.
\ee

The thermal interpretation of the model becomes transparent 
when we decompose the integrand of the partition function in modular orbits: 
\begin{itemize}
\item The $(\tilde m_0, n_0)=(0,0)$ orbit, which 
is integrated over the fundamental domain, 
\item The $(\tilde m_0,n_0)=(\tilde m_0,0)$ orbit (with $\tilde m_0\ne 0$), which is integrated over the strip.
\end{itemize}
Thanks to the initial supersymmetry, the contribution of the $(\tilde m_0, n_0)$ $=(0,0)$ vanishes.
For the latter orbit, where we set $n_0=0$, the integrand is given by 
\be
\frac{R_0}{\sqrt{\tau_2}}\, \sum_{\tilde m_0  } e^{-\frac{\pi R_0^2}{\tau_2} \tilde m_0 ^2 } \, (-1)^{\tilde m_0 (a+\bar a)}
\, \sum_{m_1 , n_1} q^{\frac{1}{4} p_{ L}^2 } \, \bar q ^{\frac{1}{4} p_{ R}^2}\, 
(-1)^{n_1 \bar b}\, e^{2i\pi \tilde m_0 \left[ G Q_+ - B Q_-\right]} \,,
\ee
where we have denoted with
$$
Q_+ = m_1 - {\textstyle\frac{1}{2}} (\bar a + n_1)\equiv {\textstyle\frac{1}{2}} R_1 (p_{L} + p_{ R} ) \,, \qquad 
Q_- = n_1 \equiv {\textstyle\frac{1}{2}} R_{1}^{-1} ( p_{ L} - p_{ R} )
$$
the $U(1)$ charges associated to the graviphoton $G_{01}$ and axial vector $B_{01}$, respectively. 
As a result, the integral over the strip is nothing but  the one-loop thermal partition function for the ${\cal N}=(4,0)$ supersymmetric model compactified on $S^1 (R_1) \times T^8$. In addition, chemical potentials for the $U(1)$ charges $Q_{\pm}$ have been turned on. These chemical potentials are imaginary. In the Euclidean context, they arise in the presence of background, classical electrostatic potentials for the $U(1)$ graviphoton and axial vector fields, or when we insert Wilson lines 
across the $x^0$ cycle for the time-like components of these fields \cite{Roberge:1986mm} \cite{KorthalsAltes:1999cp}. 
The Wilson lines are actually analogous to the Polyakov loops of finite-temperature gauge theories.

The complete space-time partition function is then given by
\begin{equation}
\label{chemical}
{\cal Z} (\beta , G, B ) = {\rm tr}\, e^{-\beta H} \, e^{2 i \pi (G Q_+ - B Q_- )} \,,
\end{equation}
where $\beta$ is the inverse temperature, and the trace is over the Hilbert space of the ${\cal N} = (4,0)$ theory.
Notice that since the Hamiltonian $H$ is quadratic in the charges $Q_+$ and $Q_-$, the partition function is real. In the case of interest, $G=2B=1$, 
and thus the contribution from the chemical potentials is nothing but the right-moving fermion index
\be
e^{2 i \pi (G Q_+ - B Q_- ) } = e^{-i \pi \bar a} = (-1)^{F_{ R}} \,.
\ee
Therefore, the partition function reads
\be
{\cal Z} (\beta , 1, {\textstyle\frac{1}{2}} ) = {\rm tr}\, e^{-\beta H}\, (-1)^{F_{ R}} = {\rm Str}\, e^{-\beta H} (-1)^{F_{ L}} \,,
\ee
where in the last equality we have introduced the graded trace. This expression makes manifest the chiral supersymmetry breaking to the $F_{L}$ and $F_{R}$ fermion numbers. 

When $G=B=0$, one recovers the conventional thermal partition function of the ${\cal N} = (4,0)$ supersymmetric model that suffers from tachyonic instabilities beyond the Hagedorn temperature. As a result, the point $G=2  B=1$ is a very special point in the thermodynamic phase diagram since only at this point the model is free of tachyons, for any values of the radii, and thus no Hagedorn instabilities occur.
Although expression (\ref{chemical}) is very useful to elucidate the general structure of the thermal partition function, we find it more convenient to analyze the $G=2B= 1$ point in terms of a left-right asymmetric non-supersymmetric deformation, since the various dualities are there manifest.

Thermal partition functions with imaginary chemical potentials have been considered in the literature before,
 in the context of non-Abelian
gauge theories, and the phase structure has been related to the confining properties of these theories \cite{Roberge:1986mm}.
It is thus interesting to consider such partition functions in the string theoretic context as well, even though their
applicability for cosmology may turn out to be a subtle issue. 
In fact, if we Fourier transform the partition function (\ref{chemical}) with respect to the periodic parameters $G$ and $B$,
we obtain a canonical thermal partition function at fixed values for the $U(1)$ charges $Q_+$ and $Q_-$:
\begin{equation}
{\cal Z} (\beta , \hat Q_+, \hat Q_- )= \Tr \left( e^{-\beta H} \delta(Q_+-\hat Q_+)\delta(Q_--\hat Q_-)\right).
\end{equation}
Therefore, the partition function at imaginary chemical potential contains physical information.

It is well known that thermal deformations, as well as deformations that lead to spontaneous breaking of supersymmetry, are equivalent to freely-acting symmetric or even asymmetric orbifolds, which combine a symmetry-breaking action with shifts along some compact cycle \cite{ShifftedLat,OpenSS}. The shifts along the compact directions make the symmetry breaking spontaneous, where the breaking scale is set by the (inverse) size of the compact cycle. In this language, the type II thermal partition function is obtained by modding out 
with $(-1)^{F_L+F_R} \, \delta$, where $\delta$ is an order-two shift along the compact Euclidean time direction.  
In the case at hand, the deformation consists of a $Z_2 \times Z_2'$ orbifold where the two {\it asymmetric} generators are
\begin{equation}
\label{generators}
g=(-1)^{F_{L}}\, \delta_0\, \in Z_2 \qquad {\rm and} \qquad 
g'=(-1)^{F_{R}}\, \delta_1\, \in Z_2' \,,
\end{equation}
with $\delta_{0,1}$ order-two shifts along the two $x^{0,1}$ cycles. The associated modular invariant partition function is not unique in this case and depends on a discrete torsion coefficient, $\epsilon =\pm 1$, in front of the disconnected twisted orbit. This discrete torsion affects the spectrum of the twisted sectors, and in particular in the $gg'$ sector it can affect the Kaluza-Klein and winding excitations of the NS-NS vacuum so that it is lifted as in eq. (\ref{tachyonmass}). 

The discrete torsion has a suggestive geometrical interpretation in terms of non trivial discrete holonomy for the NS-NS $B$-field, and indeed the non-tachyonic model actually involves $2B = 1$, which corresponds to the choice $\epsilon =-1$, as appears in the discussion above.

The interpretation of the model as a standard temperature deformation of an ${\cal N} = (4,0)$ model is equally transparent in the orbifold language, inserting in the trace the  $Z_2 \times Z_2'$ orbifold projector
\be
{\cal P} = \frac{(1 + g ) (1+ g')}{4} = 
\frac{(1 + gg' ) (1 + g')}{4} \,,
\ee
where the last equality holds since both $g$ and $g'$ are order-two elements. It is then clear that the projector $\frac{1}{2} (1 + g')$ breaks 
${\cal N} = (4,4) \to {\cal N} = (4,0)$ spontaneously along the $x^1$ cycle, while the projector $\frac{1}{2} (1 + gg')$ is a standard finite-temperature effect with non-vanishing graviphoton and magnetic fluxes.

\subsection{Touring the moduli space}

We have described a thermal model in which we have taken the $x^0$ cycle of the torus $T^2$ to be the one associated with the compact Euclidean time direction. Very similar expressions can be obtained when both the $x^0$ and $x^1$ coordinates are taken to describe spatial compact directions. However, as we already remarked at the beginning of this section, the $T^2$ moduli in this latter case correspond to propagating fields, and the interpretation of the model is dramatically different.  
In fact, these models are unstable.
To see this, it is sufficient to consider the original asymmetric model in the field direction $R_0=R_1=R$, with a diagonal metric and arbitrary $B$-field in the range $0\le 2B\le1$.  In this case, the left- and right-moving momenta in the $O \, {\bar O}$ sector read
\begin{equation}
p^0_{ L,R} = \frac{m_0 +\frac{1}{2} + B(2 n_1 +1)}{R} \pm (2 n_0 +1) R\,,
\quad
p^1_{L,R} = \frac{m_1 +\frac{1}{2} - B (2 n_0 +1)}{R} \pm (2 n_1 +1) R\,.
\end{equation}
The level-matching condition for the ground state
\begin{equation}
{\textstyle\frac{1}{4}} (p^2_{ L} - p^2_{ R})  =  \left( m_0 + {\textstyle\frac{1}{2}} \right) (2 n_0 +1) + \left( m_1 +{\textstyle\frac{1}{2}} \right) (2 n_1 +1) =0
\end{equation}
is not affected by the presence of the $B$ field, while the mass-formula (\ref{tachyonmass}) becomes
\be
m^2 _{ O \bar O} = \left( \frac{1}{\sqrt{2}\, R} - \sqrt{2}\, R\right)^2 \pm \frac{2 B(1-B)}{R^2} 
\,,
\ee
and reveals the emergence of tachyonic states whenever $B\not = 0$. 
However, the fluctuations $\delta G_{01}$ and $\delta B_{01}$ that can lead to tachyonic instabilities, in the case where both $x^0$ and $x^1$ are taken to be spatial compact directions, could be eliminated upon further (asymmetric) orbifold or orientifold projections, as in \cite{ADS}. Such tachyon free models will be studied elsewhere. 

On the other hand, if $x^0$ is taken to be the Euclidean time direction, the moduli $G_{01}$ and $B_{01}$ are frozen, and they do not correspond to fluctuating fields. This is clear in the light-cone quantization, where all oscillators along the $X^\pm$ directions are set to zero. As a result, the moduli space of the thermally deformed model consists of several disconnected components, each corresponding to different choices for the background values $G$ and $B$, having an interpretation in terms of chemical potentials. 

\section{Adding O-planes and D-branes}

The closed string sector of type II superstring theories does not by itself lead to interesting phenomenological models, since for example in typical compactifications, the gauge fields are Abelian and it does not contain matter that could describe the Standard Model particle content and its interactions. A possible way to bypass this obstruction is to construct an orientifold \cite{cargese,Polchinski,review}, where matter is localized on lower dimensional hyper-surfaces (e.g. D-branes), which can describe semi-realistically the particle content at low energies and can lead to interesting thermal cosmological models. It is thus interesting to consider the proposed tachyon free model and examine the possibility of introducing open string matter sectors via orientifold constructions. It is also challenging to obtain orientifolds starting from asymmetric realizations of spontaneous supersymmetry breaking, and studying the thermal interpretation of the matter sector.

Inspection of the closed sector partition function (\ref{part}) indicates the $Z_2$ symmetry that can be used to define the orientifold projection. Clearly, the amplitude is not invariant under the bare world-sheet parity $\Omega$. However, the combination of $\Omega$ with the permutation $\sigma: x^0 \leftrightarrow x^1$  leaves the partition function invariant, if the radii of the two compact cycles are taken to be equal.  Therefore, in this special case, 
the $Z_2$ generator $\Omega\, \sigma$ can be used to construct the orientifold.

Under the action of $Z_2$, the transformation properties of the zero modes are not conventional. In the usual orientifold 
constructions, the $B$-field is projected out from the spectrum. Here, however, due to the action of the permutation, the $B$-field remains.  Also, the field $R_0^2 - R_1^2$ is odd under the $Z_2$ symmetry
\begin{equation}
Z_2~:~\left( \psi^0_{-\frac{1}{2}} \, \tilde \psi^0_{-\frac{1}{2}} - \psi^1_{-\frac{1}{2}} \tilde \psi^1_{-\frac{1}{2}} \right) |0,\tilde 0\rangle \longrightarrow   - \left( \psi^0_{-\frac{1}{2}} \, \tilde \psi^0_{-\frac{1}{2}} - \psi^1_{-\frac{1}{2}} \tilde \psi^1_{-\frac{1}{2}} \right) |0 ,\tilde 0\rangle \,,
\end{equation}
and thus it is projected out from the spectrum.   

We are well accustomed to the fact that in conventional orientifold projections, eventually combined with T-dualities, the states that contribute to the Klein-bottle amplitude have vanishing winding or vanishing Kaluza-Klein momentum quantum numbers \cite{Polchinski,review}. The proposed orientifold is different since
\begin{equation}
Z_2 :\qquad  p_{ L}^0 = p^1_{R} \,, \qquad p_{ L}^1 = p^0_{R} \,.
\end{equation}
That is, $m_0 = m_1$ and $n_0 = - n_1$.
These conditions imply that the $Z_2$ invariant states that flow in the Klein-bottle amplitude ${\cal K}$ carry the same Kaluza-Klein and opposite winding numbers. 
The zero-mode contribution from the two-dimensional lattice to ${\cal K}$ takes the form:
\begin{equation}
 \sum_{m_0,n_0} \sum_{m_1 , n_1} q^{\frac{1}{4} \left((p^0_{\rm L})^2 + (p^1_{ L})^2 \right)} \, \bar
q^{\frac{1}{4} \left((p^0_{ R})^2 + (p^1_{ R})^2 \right)} ~ \delta_{m_0,m_1}~ \delta_{n_0 , -n_1} ~= 
~\sum_{m,n} q^{\left( \frac{m}{R} \right)^2 + (n R)^2} \,,
\end{equation}
where, as usual, $q=e^{-2 \pi \tau_2}$ in ${\cal K}$. Taking this into account, the Klein-bottle amplitude reads
\begin{eqnarray}
 {\cal K} &=& \frac{1}{2}\,\int_0^\infty \frac{d\tau_2}{\tau_2}\, \frac{P_{(8)}}{\eta^8} \, \sum_{m,n}\, \left\{ 
V_8 \, q^{\left( \frac{m}{R}\right)^2 + (2 n R)^2}
 + O_8 \, q^{\left( \frac{m+1/2}{R}\right)^2 + \left((2 n +1)R\right)^2}\right.
\nonumber \\
 && \qquad\qquad \qquad\qquad \qquad\left. 
 - S_8 \, q^{\left( \frac{m+1/2}{R}\right)^2 + (2 n R)^2}
- C_8 \, q^{\left( \frac{m}{R}\right)^2 + \left((2 n+1) R\right)^2}
\right\} \,, \label{klein}
\end{eqnarray}
where $P_{(8)}$ denotes the contributions of the momenta of the eight-dimensional lattice. Equation (\ref{klein}) represents a proper symmetrization of the torus amplitude (\ref{part}), while the low-lying spectrum depends on the interpretation of the $x^0$ coordinate. If $x^0$ labels a compact space-like coordinate then the massless spectrum comprises the eight-dimensional metric and dilaton, the single radius of the internal coordinate, $G_{01}$, the scalar $B_{01}$ together with the combinations of graviphotons $G_{\mu 1} + G_{\mu 0}$ and $B_{\mu 0} - B_{\mu 1}$, for a total of $36$ bosonic degrees of freedom. However, as we already stressed in section $2$, if $x^0$ is identified with the Euclidean time, then the light excitations comprise  the full metric and the dilaton field. 

The open-string sector is encoded in the annulus amplitude \cite{Polchinski,review}
\begin{eqnarray}
 {\cal A} &=& 
\frac{N^2}{2} \, \int_0^\infty \frac{d\tau_2}{\tau_2}\, \frac{P_{(8)}}{\eta^8}\,\sum_{m,n}\,\left\{
V_8\, q^{\frac{1}{2}  \left( \frac{m}{\sqrt{2}R}\right)^2 + \frac{1}{2} (n \sqrt{2}R)^2 }
+ O_8 \, q^{\frac{1}{2} \left( \frac{m+1/2}{\sqrt{2} R}\right)^2 + \frac{1}{2}  \left((n+\frac{1}{2}) \sqrt{2} R\right)^2 } \right.
\nonumber \\
&&\qquad\qquad \left.- S_8  \, q^{\frac{1}{2}  \left( \frac{m}{\sqrt{2} R}\right)^2 + \frac{1}{2}  \left((n+\frac{1}{2}) \sqrt{2} R\right)^2 }
- C_8\, q^{\frac{1}{2} \left( \frac{m+1/2}{\sqrt{2} R}\right)^2 + \frac{1}{2} \left(n \sqrt{2}R\right)^2 }
\right\} \,,  \label{annulus}
\end{eqnarray}
and in the M\"obius-strip amplitude
\begin{eqnarray}
 {\cal M} \!\! &=& \!\! - \frac{N}{2} \, \int_0^\infty \frac{d\tau_2}{\tau_2}\,  \frac{P_{(8)}}{\hat\eta^8}\,\sum_{m,n}\, \left\{ \hat V_8 \, q^{\frac{1}{2}  \left( \frac{m}{\sqrt{2}R}\right)^2 + \frac{1}{2}(n \sqrt{2}R)^2 }
- \hat O_8 \, (-1)^{m+n} \, q^{\frac{1}{2} \left( \frac{m+1/2}{\sqrt{2} R}\right)^2 +\frac{1}{2} \left((n+\frac{1}{2}) \sqrt{2} R\right)^2 } \right.
\nonumber \\ 
&& \qquad\qquad\left. 
- \hat S_8 \, (-1)^n \, q^{\frac{1}{2}  \left( \frac{m}{\sqrt{2} R}\right)^2 + \frac{1}{2}  \left((n+\frac{1}{2}) \sqrt{2} R\right)^2}
- \hat C_8 \, (-1)^m 
\, q^{\frac{1}{2} \left( \frac{m+1/2}{\sqrt{2} R}\right)^2 + \frac{1}{2} (n \sqrt{2}R)^2}
\right\} \,. \label{moebius}
\end{eqnarray}
Notice the presence of all characters in the open-string sector despite of the presence of a single Chan-Paton label, clearly indicating that only one type of D-brane has been introduced in this orientifold. The vectors are the only massless  states and transform in the adjoint representation of the gauge group $SO(N)$. The would-be open-string tachyon in the $O_8$ sector carries non vanishing momentum and winding numbers and therefore it is always massive at generic values of the $R$-modulus, while at the fermionic point it contributes with some additional massless states that, as we will see later on, transform in the symmetric and antisymmetric representations of $SO(N)$.

As usual, the rank of the gauge group is determined by tadpole cancellation conditions for massless states, extracted from of  the transverse closed-string channel \cite{Polchinski,review}:
\begin{eqnarray}
 2 {\cal \tilde K} &=& 2^4 \, vol_8\, \int_0^\infty d\ell ~ \frac{W_{(8)}}{\eta^8}~ \sum_{m,n}\, \left[ 
V_8 \, q^{\frac{1}{2}\left( \frac{2m}{R}\right)^2 +\frac{1}{2} ( n R)^2}
 + O_8 \, q^{\frac{1}{2}\left( \frac{2m+1}{R}\right)^2 +\frac{1}{2} \left(( n +\frac{1}{2})R\right)^2} \right.
\nonumber \\
&& \left.  - S_8 \, q^{\frac{1}{2} \left( \frac{2m+1}{R}\right)^2 + \frac{1}{2}( n R)^2}
- C_8 \, q^{\frac{1}{2}\left( \frac{2m}{R}\right)^2 +\frac{1}{2} \left(( n+\frac{1}{2}) R\right)^2} \right] \,,
\label{Ktilde}
\end{eqnarray}
\begin{eqnarray}
2 {\cal \tilde A} &=&\frac{N^2}{2^4}\, vol_8 \, \int_0^\infty d\ell ~\frac{W_{(8)}}{\eta^8}~ \sum_{m,n}\, \left[
V_8\, q^{\frac{1}{2} \left( \frac{2m}{R}\right)^2 + \frac{1}{2}  (n R)^2 }
+ O_8 \, q^{\frac{1}{2}  \left( \frac{2m+1}{ R}\right)^2 + \frac{1}{2} \left((n+\frac{1}{2}) R\right)^2 } 
\right.
\nonumber \\
&& \left. - S_8  \, q^{ \frac{1}{2} \left( \frac{2m+1}{ R} \right)^2 + \frac{1}{2}  (n R)^2 }
- C_8\, q^{\frac{1}{2} \left( \frac{2m}{R}\right)^2 + \frac{1}{2}\left( (n +\frac{1}{2} ) R\right)^2 }\right] \,, 
\label{Atilde}
\end{eqnarray}
and
\begin{eqnarray}
 {2 \cal \tilde M}  &=& -2N~ vol_8\, \int_0^\infty d\ell ~ \frac{W_{(8)}}{\eta^8}~ 
 \sum_{m,n}\,  \left[ \hat V_8 \, 
 q^{ \frac{1}{2} \left( \frac{2m}{R}\right)^2 + \frac{1}{2} (n R)^2 }
+ \hat O_8 \, (-1)^{m+n} \, q^{\frac{1}{2} \left( \frac{2m+1}{ R}\right)^2 +\frac{1}{2}  \left((n+\frac{1}{2})  R\right)^2 }  \right.
\nonumber \\
&&\qquad \qquad \left. - \hat S_8 \, (-1)^m \,q^{\frac{1}{2} \left( \frac{2m+1}{ R}\right)^2 +\frac{1}{2}  \left(n R\right)^2 } 
- \hat C_8 \, (-1)^n \, q^{\frac{1}{2} \left( \frac{2m}{ R}\right)^2 +\frac{1}{2}  \left((n+\frac{1}{2})  R\right)^2 }  \right]
 \,,
\end{eqnarray}
 where $vol_8$ denotes the volume of the eight-dimensional space, and $W_{(8)}$ includes the contribution of the associated closed-string winding modes.
 Clearly, this model is free of RR tadpoles since in the deformed type IIB model (\ref{part}), the RR sectors are massive. The dilaton tadpole reads
\begin{equation}
\label{vvtadpole}
\tilde{\cal K} + \tilde {\cal A} + \tilde{\cal M}~~~ {\rm for}~~ V_8 :~ 2^4 + 2^{-4} \, N^2 -2\, N =0 \,,
\end{equation}
and therefore fixes the gauge group to be $SO(16)$. We should stress here that the dilaton tadpole is always present at generic points of the moduli space, and can be canceled by (\ref{vvtadpole}).  

At the special fermionic point, extra massless states arise from the $O_8$ sector that require 
an additional tadpole cancellation condition
\begin{equation}
\tilde{\cal K} + \tilde {\cal A} + \tilde{\cal M}~~~ {\rm for}~~O_8:  2^4 + 2^{-4}\, N^2 - 2 \, N \, \sum_{m,n=-1,0}(-1)^{m+n} ~=~ 2^4 + 2^{-4} N^2 \ne 0 \,. \label{newtadpole}
\end{equation}
This cannot be satisfied since the term proportional to $(-)^{m+n}$ vanishes. We stress here that this tadpole occurs only at the fermionic point. For cosmological considerations, we would like to include the fermionic point, and furthermore use it to define the early phase of the cosmology. Thus, it is crucial to seek other consistent projections involving a different choice of the M\"obius amplitude, leading to a cancellation of this tadpole. 
Had we left the tadpole uncanceled, the $SO(4)$ gauge symmetry enhancement at the fermionic point would be spoiled, since 
the relevant massless states from the $O_8$ sector are charged under this. The problem is similar to the one occurring due to massless RR tadpoles in standard orientifolds.

A consistent choice for the  M\"obius amplitude which respects the $SO(4)$ gauge symmetry
at the fermionic point is the following: 
\begin{eqnarray}
{\cal M}' &=& -\frac{N}{2} \, \int_0^\infty \frac{d \tau_2}{\tau_2} \, \frac{P_{(8)}}{\eta^8}\, \sum_{m,n} \left\{
\left( \hat V_6 \,\hat O_2 - \hat O_6 \, \hat V_2 \right) \, (-1)^{m+n} \,  q^{\frac{1}{2}  \left( \frac{m}{\sqrt{2}R}\right)^2 + \frac{1}{2} (n \sqrt{2}R)^2 } 
\right.
\nonumber \\
&& \qquad\qquad \qquad\qquad 
~~~ - \left( \hat O_6 \, \hat O_2 + \hat V_6 \, \hat V_2 \right) \, 
 q^{\frac{1}{2} \left( \frac{m+\frac{1}{2}}{\sqrt{2}R}\right)^2 +\frac{1}{2}  \left((n+\frac{1}{2}) \sqrt{2}R\right)^2 }
 \nonumber \\
& &~~~ \qquad\qquad \qquad\qquad
- \left( \hat S_6 \,\hat S_2 + \hat C_6 \, \hat C_2 
\right)\, (-1)^n \, q^{\frac{1}{2} \left( \frac{m}{\sqrt{2} R}\right)^2 +\frac{1}{2}  \left((n+\frac{1}{2}) \sqrt{2} R\right)^2 }
\nonumber \\
& &~~~\qquad\qquad \qquad\qquad \left.
- \left( \hat S_6 \, \hat C_2 + \hat C_6 \, \hat S_2  
\right) \, (-1)^m
\, q^{\frac{1}{2} \left( \frac{m+1/2}{\sqrt{2} R}\right)^2 + \frac{1}{2} \left(n \sqrt{2}R\right)^2}
\right\},
\end{eqnarray}
where we have decomposed  the $SO(8)$ characters in terms of $SO(6)\times SO(2)$ ones,  since after all $SO(6)\times SO(2)$ is the residual symmetry group of this asymmetric model.
Now, the massless space-time vectors  coming from  the sector $  V_6 \, O_2 $,  with multiplicity $N(N-1)/2$, are in the antisymmetric (adjoint) representation of the gauge group $SO(16)$, while the two massless scalars 
coming from  the sector $  O_6 \, V_2 $,  with multiplicity $N(N+1)/2$, are in the  {\bf 136}-dimensional symmetric representation of  $SO(16)$, due to the change of sign of the corresponding term in  ${\cal M}' $. The four extra massless scalars at the fermionic point coming 
from the sector  $O_6\, O_2$,   with multiplicity $N(N+1)/2$,  are also in the symmetric representation of  $SO(16)$.  Therefore, at the fermionic point,  where an enhanced $SO(4)$ gauge symmetry emerges from the closed string sector,  the $6 \times N(N+1)/2$ massless scalars from the  $  O_6 \, V_2 $  and $O_6\, O_2$ sectors are simultaneously charged under the gauge groups from the closed and open string sectors. In particular, they are in the representation $({\bf 6}\,,\, {\bf 136})$ of $SO(4)_{\rm closed} \times SO(16)_{\rm open}$. To the best of our knowledge, this is the first instance where open-string states carry, at the same time, non-trivial representations of Chan-Paton and closed-string non-abelian gauge groups.  

In order to analyze the tadpole cancellation we need the 
expression  of $\cal \tilde M'$ in the transverse channel:
\begin{eqnarray}
2{\cal \tilde M'}&=&2N~vol_8\, \int_0^\infty d\ell \, \frac{W_{(8)}}{\eta^8}~ \sum_{m,n}~\left[ 
\left( \hat V_6 \, \hat O_2 - \hat O_6 \, \hat V_2 \right)
~ (-1)^{m+n}~ q^{ \frac{1}{2} \left( \frac{2m}{R}\right)^2 + \frac{1}{2} (n R)^2 } \right.
\nonumber \\
&&  \quad\qquad \qquad \qquad\qquad -\left( \hat O_6 \, \hat O_2 +\hat V_6 \, \hat V_2 \right) 
~q^{\frac{1}{2} \left( \frac{2m+1}{ R}\right)^2 +\frac{1}{2}  \left[(n+\frac{1}{2})  R\right]^2 }  
\nonumber \\
&& \quad\qquad \qquad \qquad\qquad- \left( \hat S_6 \, \hat S_2 + \hat C_6 \, \hat C_2 \right) \, (-1)^m \,q^{\frac{1}{2} \left( \frac{2m+1}{ R}\right)^2 +\frac{1}{2}  \left(n R\right)^2 } 
\nonumber \\
&& \quad \qquad \qquad \qquad\qquad\left. - \left( \hat C_6 \, \hat S_2 + \hat S_6 \, \hat C_2 \right) \, (-1)^n \, q^{\frac{1}{2} \left( \frac{2m}{ R}\right)^2 +\frac{1}{2}  \left((n+\frac{1}{2})  R\right)^2 }  \right]
\,.
\end{eqnarray}
The dilaton tadpole in $V_6 O_2$ is not canceled now,
\be
\tilde{\cal K} + \tilde {\cal A} + \tilde{\cal M'}~~~ {\rm for}~~ V_6O_2 :~ 2^4 + 2^{-4} \, N^2 +2\, N\ne 0, 
\ee
 while the tadpole for the would-be tachyon $O_6 O_2$ and for the 
internal metric components $O_6 V_2$, forming a full representation of the  
$SO(4)$ symmetry, are both canceled: 
\be
\tilde{\cal K} + \tilde {\cal A} + \tilde{\cal M'}~~~ {\rm for}~~ O_6 O_2+ O_6 V_2 :~ 2^4 + 2^{-4} \, N^2 -2\, N =0~~{\rm when~~}N=16.
\ee
Thus as a consequence of the change of signs in the M\"obius amplitude $\cal M'$, the tadpoles for the ``charged'' $SO(4)$ closed-string states are now canceled, at the cost of a non-cancellation of the dilaton tadpole. The latter however is harmless and indicates that the flat Minkowski vacuum is unstable.

Notice that to map the direct-channel and transverse-channel M\"obius amplitudes one must use the $P = T^{1/2}S T^2 S T^{1/2}$ transformation that on the $SO(6)$ and $SO(2)$ characters acts like \cite{review} 
\begin{eqnarray}
&& \hat O_6 \rightarrow \frac{1}{\sqrt{2}}\left( -\hat O_6+\hat V_6  \right),
~~~~~~~~~\hat O_2 \rightarrow \frac{1}{\sqrt{2}}\left( \hat O_6+\hat V_2  \right),
\nonumber \\
&&\hat V_6 \rightarrow \frac{1}{\sqrt{2}}\left(~~ \hat O_6+\hat V_6  \right),~~~~~~~~~~\hat V_2 \rightarrow \frac{1}{\sqrt{2}}\left( \hat O_2-\hat V_2  \right).~
\end{eqnarray}

It is interesting to give a thermal interpretation of the amplitudes (\ref{annulus}) and (\ref{moebius}) (when $x^0$ is interpreted as the Euclidean time). To start with, we note that the amplitude is T-duality invariant under the transformation $\sqrt{2} R\to 1/\sqrt{2} R$,  reversing also the chirality of the space-time spinors. Moreover, the Ramond sectors $S$ and $C$ have dual masses while the $O$ sector is self-dual and massive. Therefore, on this open-string sector the $R$-modulus deformation acts precisely as a self-dual thermal deformation, consistent with the low and high temperature limits $\beta \equiv R\to \infty$ and $\beta \equiv R\to 0$ of (\ref{annulus}) and (\ref{moebius}). In the former case, the $V$ and $S$ sectors survive while the others become infinitely massive and decouple from the spectrum. In the latter case, the $V$ and $C$ sectors survive while the others become infinitely massive and decouple. In either case, the would be tachyons in the $O$ sector decouple, together with one  spinor representation, and supersymmetry is recovered.

Before we conclude this section, let us try to give a geometrical interpretation of the O-planes and D-branes involved by this asymmetric action. Clearly, the orientifold planes are localized at the fixed points of the orientifold projector, $\Omega \sigma$ \cite{Polchinski,review}. As a result, we have O8 planes stretched along the diagonals of the $T^2$, since the $x^0 \pm x^1=0$ is invariant under $\sigma$, modulo identifications on the lattice. Moreover, from eq. (\ref{Ktilde}) one can determine that they have the same NSNS tension, while they carry opposite RR charges, and have opposite couplings to the would be tachyon, so yielding together a neutral, though massive, configuration. This is actually consistent with the fact that, in the closed-string sector, there are no massless RR fields they can couple to. Notice that the O8 planes here involved are suitable deformations of those that appear in type 0B orientifolds \cite{review}, that in general have non-vanishing couplings to all NSNS and RR sectors.

Turning to the D-branes, as already anticipated before, they have the peculiarity that all sectors are simultaneously present on a given stack. This has to be contrasted to the conventional BPS D-branes, where only the supersymmetric combination $V_8 - S_8$ is present, and indeed correspond to the non-BPS branes of Sen \cite{Sen}. They can be thought of as bound states of branes and anti-branes \cite{DMS} and therefore carry no charge with respect to the RR fields. Clearly the asymmetric ``thermal'' deformations act also on the non-BPS D8-branes yielding non-trivial masses to the fermions and the would-be open-string tachyon. Moreover, from the zero-modes contributions to ${\cal A}$, one can immediately deduce that also these D8-branes are stretched along the diagonal, whose length is indeed $\sqrt{2} \, R$.

\section{Conclusions and perspectives}
In this work we study type II superstring models where supersymmetry is spontaneously broken via asymmetric gravito-magnetic fluxes. In the closed string sector, the partition function is shown to be free of Hagedorn instabilities for a certain choice of the fluxes. The models describe thermal ensembles which are deformed by chemical potentials. All fermions  acquire masses, as it is the case in finite temperature systems, however here the RR bosonic states also acquire masses due to the gravito-magnetic fluxes. 
In the open string  matter sector, the free energy is qualitatively similar to the conventional thermal free energy in that the initially massless bosonic spectrum is unaffected by the deformation. In both sectors the free energy satisfies a novel Temperature duality ${\cal F}(T/T_H)={T^2 \over T_H^2}~{\cal F}(T_H/T)$. 
As the temperature approaches the Hagedorn temperature, the behavior of the system deviates drastically from the conventional thermal one, where now no tachyons are generated above $T_H$ due to the Temperature duality. 
At the self-dual point $T=T_H$, extra bosonic states become massless. This point is characterized by enhanced non-abelian $SO(4)$ gauge symmetry. 

Changing the temperature adiabatically suggests the existence of a non-singular phase of the Universe where the $SO(4)$ gauge symmetry is restored. Since the free energy is non-trivial and finite, the back-reaction on the initially flat background will presumably induce a cosmological evolution. For small temperatures this evolution is very similar to that of a radiation dominated expanding Universe \cite{Cosmo-RT-Shifted, Mod-Stab}. 
For relatively late times the corresponding cosmology is under investigation.             
The cosmological backreaction around the self-dual point is very interesting to examine, and hopefully it will shed some light on the structure of the early universe. 
We expect that this structure is characterized by a stringy, non-geometrical phase. Following the lines of \cite{massiveSUSY}, this suggests the absence of space-like singularities in the cosmology. 

Furthermore, following references \cite{Kutasov:1990sv,Israel:2007nj} the absence of Hagedorn-like singularities for any $T$ suggests an asymptotic spectrum degeneracy of bosonic and fermionic massive string states. This can be easily shown to be valid using the properties of the $SO(8)$ characters and their couplings to the $\Gamma_{(2,2)}$ shifted lattice.

\vskip 1.0cm

\section*{Acknowledgements}

We are  grateful to Luis Alvarez-Gaum\'e, Joseph Polchinski, Giovanni Ricco and especially  to Manolis Floratos and Jan Troost for useful and fruitful  discussions. C.A. and N.T. would like to thank the Laboratoire de Physique Theorique of Ecole Normale Superieure and the Centre de Physique Theorique of Ecole Polytechnique for hospitality. 
C.K. would like to thank the University of Cyprus for hospitality. H.P. would like to thank the Laboratoire de Physique Theorique of Ecole Normale Superieure and  the University of Cyprus for hospitality. 
The work of C.A., C.K. and H.P. is partially supported by the EU contract MRTN-CT-2004-005104 and the ANR (CNRS-USAR) contract  05-BLAN-0079-01. 
 C.K and N.T.  are supported by the EU
contract MRTN-CT-2004-512194. 
 C.A. is also partially supported by the Italian MIUR-PRIN contract 20075ATT78.
 H.P. is also supported by the EU contracts MRTN-CT-2004-503369 and
MEXT-CT-2003-509661, INTAS grant 03-51-6346, and CNRS PICS 2530, 3059 and 3747.
N.T. is also supported by an INTERREG IIIA Crete/Cyprus program.

\end{document}